# Problems with modelling closed timelike curves as post-selected teleportation

T.C. Ralph

*Centre for Quantum Computation and Communication Technology,*
*School of Mathematics and Physics, University of Queensland, St Lucia, Queensland 4072, Australia*
(Dated: November 19, 2018)

In this comment on S.Lloyd, *et al*, Phys.Rev.Lett. **106**, 040403 (2011), we show that modelling closed timelike curves (CTCs) as post-selected teleportation allows signalling to past times *before* the creation of the CTC and allows information paradoxes to form.

In their interesting recent letter Lloyd et al [1] propose post-selected teleportation as an alternative model for quantum systems interacting with closed timelike curves (CTCs) - paths through spacetime, apparently allowed by general relativity, that allow objects to interact with their own past - contrasting it with the more accepted model due to Deutsch [2]. The authors suggest that their post-selected model (P-CTC) is able to avoid certain paradoxes associated with time travel whilst retaining all correlations of the interacting quantum system with other quantum systems. Ref [2] does not allow quantum correlations with other systems to be retained in general. Here we show by example that when entanglement is considered, P-CTCs: (i) allow signalling to past times *before* the creation of the closed timelike curve; and (ii) allow an unresolved "unproven theorem" paradox to form in this epoch.

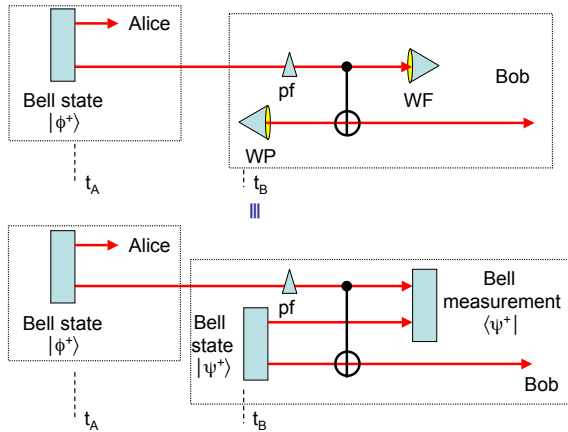

FIG. 1: Top panel: model of entanglement interacting with a closed timelike curve formed by a wormhole. WF is the future mouth of the wormhole; WP is the past mouth of the wormhole; pf is a phase flip that may be applied. The interaction is a CNOT gate. Bottom panel: according to [1] this is equivalent to the depicted post-selected teleportation circuit in which the input state is renormalized to the post-selected outcome of the Bell measurement.

We consider passing one arm of an entangled state through a time-machine wormhole [3]. In particular let us suppose that the entangled state $|\phi^+\rangle = 1/\sqrt{2}(|0\rangle_1|1\rangle_2 + |1\rangle_1|0\rangle_2)$ is created by Alice at some time $t_A$. She keeps one arm and sends the other to Bob. At some later time $t_B$ Bob creates a compact wormhole time-machine that enables him to send his entangled particle a short time into the past - CTCs form for times $t > t_B$. Bob chooses to perform, or not, a phase flip on his qubit before sending it through the wormhole in the manner of Fig.1 (top panel), allowing it to interact with its past self via a CNOT gate. According to [1] the output of this interaction should be obtained by considering the equivalent teleportation circuit Fig.1 (bottom panel), renormalized to only the outcome for which the same Bell-state is detected at the Bell measurement as the Bell state that drives the teleporter. Applying this approach it is straight forward to calculate the state of Alice and Bob is: $\frac{1}{\sqrt{2}}(|0\rangle \pm |1\rangle)_A|0\rangle_B$, where the sign is positive (negative) for no phase flip (when Bob applies a phase flip). By choosing to introduce a phase flip on his qubit or not, Bob can deterministically send a string of bits (in the diagonal basis) to Alice in the past. The existence of this "radio to the past" creates the possibility of new paradoxes, now in the classical domain. In particular the unproved theorem paradox [2] can arise. Bob reads the proof of a theorem in a book and uses the set-up of Fig.1 (top panel) to send it back in time to Alice. Alice subsequently publishes it in the book from which Bob will read it. Where did the proof come from?

To show clearly that this paradox is not resolved by the P-CTC approach we describe how it could be simulated with the circuit of Fig.1 (bottom panel). Alice reads out data from her half of the entangled pair in the diagonal basis and publishes it in a "book". The data will be a random bit string. The editor of the book, Oscar, manipulates the data in the following way: he identifies "message bits" in the data that he leaves alone; and he identifies "non-message bits" in the data which he bit flips. He writes the theorem proof into the data by picking out a particular sequence of message bits, and making all other bits non-message bits. Bob reads the book and sends the data to Alice by manipulating the phase flip. When we postselect only the outcomes for which the correct Bell-state is detected we find only those events corresponding to message bits are kept. According to Ref [1], we should interpret the post-selected events as the *only* events that would occur if we used an actual CTC. But there is no editing of the post-selected events, i.e. Oscar does not exist (or does nothing). The post-selected book contains only the theorem proof, directly as it ap-peared in Alice's data, which Bob dutifully sends (sent) back to her in the past. In the post-selected universe the proof appears from nowhere. The attempt to resolve acausal interactions with the near past by the CTC via the method of [1], leads to information flow arbitrarily far into the past through the entanglement and the appearance of unresolved information paradoxes.

In contrast the decorrelation of entanglement that occurs through the method of [2] ensures that acausal effects of the CTC do not spread outside the CTC epoch. Information paradoxes are resolved by Deutsch's maximum entropy conjecture, which has recently been independently derived [4].

---